\documentstyle[epsf,aps]{revtex}
\draft
\begin{document}
\title{Anisotropic Vacuum Induced Interference in Decay Channels}
\author{G.S. Agarwal\footnote{also at Jawaharlal Nehru Centre for Advanced Scientific
Research, Bangalore, India}}
\address{Physical Research Laboratory\\
 Navrangpura, Ahmedabad-380 009, INDIA}
\date{21$^{st}$ Mar. 2000}
\maketitle
\begin{abstract}
We demonstrate how the anisotropy of the vacuum of the electromagnetic field
can lead to quantum interferences among the decay channels of close lying states. 
Our key result is that interferences are given by the {\em scalar} formed from
the antinormally ordered electric field correlation tensor for the anisotropic
vacuum and the dipole matrix elements for the two transitions.
We present results for
emission between two conducting plates as well as for a two
photon process involving fluorescence produced under coherent cw excitation.
\end{abstract}

Recently considerable effort has been devoted to the question of coherence and
interference effects arising from the decay of close lying energy levels 
\cite{swain1,knight99,zhu,xia96,rus99,arwal74,seharris,scully98,paspal,menon98}. 
Such interference effects lead to many remarkable phenomena such as population 
trapping \cite{arwal74}, spectral narrowing \cite{swain1,knight99,zhu}, gain 
without inversion \cite{swain1,seharris,paspal}, phase dependent line shapes
\cite{knight99,menon98,nar97} and quantum beats \cite{rus99,pat98}. However the very existence of the interference effect
depends on the validity of a very stringent condition viz. the dipole matrix
moments for two close lying states decaying to a common final state should be
{\bf non-orthogonal} \cite{arwal74}.
This last condition is really the bottleneck in the observation of the predicted
new effects. Some progress however was made by the use of static and
electromagnetic fields to mix the levels \cite{pat98} so that the relevant dipole
moments become non-orthogonal. In this letter, we propose a totally {\it
different mechanism} to overcome the problem of the orthogonality of the dipole 
moments. We suggest working in such situations where the vacuum of the
electromagnetic field is {\it anisotropic}, so that the interference among decay
channels can occur even if the corresponding dipole moments are orthogonal. This
provides a possible solution to the long standing problem in the subject of
interference among decay channels. Our key result is that interferences are 
given by the {\it scalar} formed from the antinormally ordered electric field 
correlation tensor for the anisotropic vacuum and the dipole matrix elements 
for the two transitions.  This is in contrast to the usual result that 
interferences are given in terms of the {\em scalar} formed out of the 
dipole matrix elements.  This opens up the possibility of studying quantum
interferences in a variety of new class of systems.
At the outset, we mention that one can
consider many situations where vacuum will be anisotropic, for example (a)
doped active centers in anisotropic glasses \cite{rikken}, (b) emission of active
atoms in a waveguide \cite{brorson}, (c) spontaneous emission from atoms adsorbed on
metallic or dielectric surfaces \cite{saran75}, (d) emission in a spatially
dispersive medium - which allows the possibility of longitudinal electromagnetic
fields, (e) emission between two conducting plates \cite{saran75,knmildow}, 
which is a problem of great interest since the early work of Casimir.  
Our results suggest the study of quantum interferences in a totally new class
of situations involving atoms
and molecules adsorbed on surfaces.
Explicit results in some of these situations would be given later.

{\bf Interferences in Fluorescence under Coherent Excitation -Two 
Photon Processes} 

Before we present detailed dynamical equations, we consider a simple
situation which enables us to bring out the essential physics of the
interferences associated with anisotropic vacuum fields.   We basically examine
the nature of interferences in spontaneous emission.  However here the way the
system is excited is important.  A practical way will be to excite by a coherent
cw radiation.  In technical terms this is  a two-photon or a second order
process.  This is in contrast to those studies which ignore how the system was
excited.  Note that in the experiment of Xia et al. \cite{xia96} the fluorescence
produced by two photon excitation was studied.  
Let us consider the process[Fig. 1]
in which the atom in the state $|g\rangle$ absorbs a photon
of frequency $\omega_l$ and wavevector $\vec{k_l}$ and emits a photon to end
up in the state $|f\rangle$ which is distinct from the ground state $|g\rangle$.
In the process of absorption and emission,  the atom goes through a number of
virtual intermediate states. For the purpose of the argument, we retain only two
intermediate states $|j\rangle~~(j=1,2)$. The transition probability for this
process can be calculated using the second order perturbation theory
(cf. ref. \cite{loudon}).
The initial state of the field is vacuum $|v\rangle$. For simplicity we assume
that the absorption is from a coherent field ${\cal{E}}_l$. Let $H_I(t)$ be the
interaction Hamiltonian in the interaction picture. As usual \cite{loudon} we assume that the
perturbation is switched on slowly. Then the transition probability of the above
second order process can be written as
\begin{equation}
T_{gf}=\lim_{t\rightarrow\infty }\frac{d}{dt}\sum_F\left|\frac{1}{\hbar^2}
\int_{-\infty}^t  dt_1\int_{-\infty}^{t_1} dt_2\langle f,~ F\left| H_I(t_1)H_{I}(t_2) e^{\epsilon(t_1 + t_2)}
\right | g, v \rangle \right |^2 .
\end{equation}
We sum over all final states $|F\rangle$ of the field i.e. we assume that no spectral
measurement of the emitted field is done. The interaction Hamiltonian can be
written as 
\begin{eqnarray}
H_I (t) &=& -\vec{d}(t).\vec{\cal E}_l~ e^{-i\omega_l
t+i\vec{k}_l.\vec{r}} \nonumber \\
&-&\vec{d}(t).\vec{E}_{v} (t)+ {\rm H.C.}~,
\end{eqnarray}
where $\vec{E}_{v}$ is the electric field operator for the vacuum 
and $\vec{d}$ is the dipole moment operator for the
atom. On substituting (2) in (1) and on carrying out all the simplifications
and on making rotating wave approximation we find the expression for the
transition probability 
\begin{equation}
T_{gf} = \frac{1}{\hbar^2}\sum_{i,j}\frac{g_i g_j^* \vec{d}_{fj}^*.
\stackrel\Rightarrow{C} (\omega_l-\omega_{fg}).\vec{d}_{fi}}
{(\omega_{ig} - \omega_l)(\omega_{jg}-\omega_l)},~~~~ g_i =
\frac{\vec{d}_{ig}.\vec{\cal E}_{l}}{\hbar} .
\end{equation}
Here we have introduced the correlation function tensor
$\stackrel{\Rightarrow}C$ for the anti-normally ordered correlation function
for the vacuum field 
\begin{equation}
\stackrel{\Rightarrow}C(\omega)= \int_{-\infty}^{+\infty} dt 
~e^{i\omega t}\langle\vec{E}_v^{(+)}(t)\vec{E}_v^{(-)} (0)\rangle ,~~~~
\vec{E}_v(t)=\vec{E}_v^{(+)}(t) + \vec{E}_v^{(-)}(t),
\end{equation}
where $\vec{E}_v^{(+)}$ and $\vec{E}_v^{(-)}$ are respectively the positive and negative
frequency parts of the field operator representing anisotropic vacuum. The two field operators in (4) are to be 
evaluated at the position of the atom. The expression (3) displays explicitly
the atomic and the vacuum field characteristics. The {\it anisotropic vacuum} enters
through the {\it correlation tensor} $\stackrel{\Rightarrow}C.$
The terms $i\neq j$ in (3) correspond to the interferences between the decay
channels $|i\rangle\rightarrow|f\rangle$ and $|j\rangle\rightarrow|f\rangle$.
The quantum interferences will be non-vanishing only if
\begin{equation}
\vec{d}_{fj}^* .\stackrel{\Rightarrow}C (\omega_l - \omega_{fg}).
\vec{d}_{fi}\neq 0.
\end{equation}
This is one of the key results of this paper. For isotropic vacuum the
correlation tensor $\stackrel{\Rightarrow}C$ is proportional to the unit tensor:
$\stackrel{\Rightarrow}C =\stackrel{\Rightarrow}I C$ and hence (5) reduces to
\begin{equation}
\vec{d}_{fj}^* . \vec{d}_{fi}\neq 0.
\end{equation}
Clearly the interferences will survive even if the corresponding dipole matrix
elements are orthogonal provided that the vacuum field is anisotropic. Note further
that with a proper tuning of the field the amplitude $T_{gf}$ can, in principle,
become zero. It may be noted that the correlation functions
$\stackrel{\Rightarrow}C$ are known in the literature for a variety of 
situations including the ones mentioned in the introduction.

{\bf Anisotropy Induced Interference in Dynamical Evolution}

Let us next consider the dynamical evolution of the atomic density matrix so that we
can study various line shapes and other dynamical aspects of emission. For
simplicity we consider a $j=1$ to $j=0$ transition. Let a static magnetic
field be applied along $y$ direction. This defines the quantization axis. The
magnetic sub-level $|1\rangle\equiv|j=1,~m=1\rangle$ with energy $\hbar\omega_1$
($|2\rangle\equiv|j=1, 
m=-1\rangle$ with energy $\hbar\omega_2$) decays to the state $| 3 \rangle=|j=0,~m=0\rangle$ (energy = 0) with the
emission of a right (left) circularly polarized photon. We drop the level
$j=1,~m=0$ from our consideration as it does not participate in interferences.
The Hamiltonian for interaction of the atom with the vacuum is
\begin{eqnarray}
H_1 = &+& d| 1\rangle\langle 3 | \hat{\epsilon}_{-}.\vec{E}_{v}(t) \nonumber\\
 &-& d| 2\rangle\langle 3 |\hat{\epsilon}_+.\vec{E}_{v}(t) ~~+{\rm H.C.} ,
\end{eqnarray}
where $\hat{\epsilon}_{\pm}=(\hat{z}\pm i\hat{x})/\sqrt{2}$ and $d$ is the
reduced dipole matrix element. In order to describe the dynamics of the atom, we
use the master equation framework. We use the Born and Markov approximations to
derive the master equation for the atomic density matrix $\rho$. In rotating
wave approximation, our calculations lead to the equation
\begin{eqnarray}
\frac{\partial\rho}{\partial t}=
&-& i(\omega_1 |1\rangle 
\langle 1|+\omega_2 |2\rangle\langle 2|)\rho \nonumber \\
&-&\gamma_1 (|1\rangle\langle 1|\rho -
\rho_{11}|3\rangle\langle 3|) \nonumber \\
&-&\gamma_2 (|2\rangle\langle 2|\rho -\rho_{22}|3\rangle\langle 3|)\nonumber\\
&+& \kappa_2 (|1\rangle\langle 2|\rho -\rho_{21}|3\rangle\langle 3|)\nonumber\\
&+& \kappa_1 (|2\rangle\langle 1|\rho -\rho_{12}|3\rangle\langle 3|) +{\rm
H.C.}.
\end{eqnarray}
Here the coefficients $\gamma's$ and $\kappa's$ are related to the antinormally ordered
correlation functions of the vacuum field
\begin{equation}
\stackrel{\Rightarrow}C^{(+)}(\omega) =
\int_0^\infty\langle\vec{E}_{v}^{(+)}(\tau)\vec{E}_{v}^{(-)} (0)\rangle
~e^{i\omega\tau }d\tau,
\end{equation}
\begin{equation}
\gamma_1 = \frac{d^2}{\hbar^2}\hat{\varepsilon}_- .\stackrel{\Rightarrow}C (\omega_1).
\hat{\varepsilon}_+~~;~~
\gamma_2=\frac{d^2}{\hbar^2}\hat{\varepsilon}_+.\stackrel{\Rightarrow}C (\omega_2).\hat{\varepsilon}_{-} ,
\end{equation}
\begin{equation}
\kappa_2 =
\frac{d^2}{\hbar^2}\hat{\varepsilon}_-.\stackrel{\Rightarrow}C(\omega_2).
\hat{\varepsilon}_-~~;~~
\kappa_1 = \frac{d^2}{\hbar^2}~\hat{\varepsilon}_+.\stackrel{\Rightarrow}C (\omega_1).
\hat{\varepsilon}_+.
\end{equation}
Note that the terms involving $\kappa_1$ and $\kappa_2$ are responsible for
interferences between the two decay channels $|1\rangle\rightarrow |3\rangle$
and $|2\rangle\rightarrow |3\rangle$. For the case of free space, 
vacuum is isotropic 
\begin{equation}
C_{zz}^{(+)} (\omega) = C_{xx}^{(+)}(\omega),~C_{zx}^{(+)}
(\omega)=C_{xz}^{(+)} (\omega) = 0 ~,
\end{equation}
and hence
\begin{equation}
\kappa_1 = \kappa_2 = 0,
\end{equation}
leading to no interferences in the decay channels. Clearly,  for decay in free
space the interferences could be possible only if the dipole matrix elements
were non-orthogonal:  $\vec{d}_{13}.\vec{d}_{23}^* \neq 0$. Our development of the
master equation shows how the interferences in the decay channels are possible
even if the dipole matrix elements were orthogonal. We need the {\it anisotropy}
of the vacuum. The anisotropy leads to the non-vanishing of the coefficients
$\kappa_1$ and $\kappa_2$.  The interferences are particularly prominent when
$\kappa$'s become comparable to $\gamma$'s.  Thus for our situation we will
recover all previous results \cite{swain1,knight99,zhu,xia96,rus99,arwal74,seharris,scully98,paspal,menon98,nar97} on line shapes and trapping.

We could now consider explicitly situations of the type mentioned in the 
introduction. The correlation functions $\stackrel{\Rightarrow}C$
can be computed for example in situations corresponding to the emission from an
atom in a metallic waveguide or an atom between the plates of a perfect
conductor. On using the relation 
\begin{equation}
\frac{1}{{\varepsilon} + i(\omega_0-\omega)} = P\frac{1}{i(\omega_0-\omega)}
+ \pi\delta(\omega_0-\omega) ,
\end{equation}
and on ignoring the principal value terms in Eq.(9), we can approximate
$\stackrel{\Rightarrow}C^{(+)}$ as
\begin{equation}
\stackrel{\Rightarrow}C^{(+)}(\omega)\approx\frac{1}{2}\stackrel{\Rightarrow}C(\omega).
\end{equation}

The correlation function for the vacuum field can be calculated by quantizing
the field and by using the properties of annihilation and creation operators.
However in certain situations the explicit quantization of the field is
complicated and hence we follow a {\em different} method.
Using the {\em linear response theory} the correlation function 
$\stackrel{\Rightarrow}C(\omega)$ can be related \cite{agar75}  to the
solution $\vec{E}(\vec{r},t)=\vec{\cal E}(\bar{r},\omega) e^{-i\omega t} +{\rm
c.c.}$  of  Maxwell's  equations with a source polarization  $\vec{P}\equiv\vec{p}\delta~
(\vec{r}-\vec{r_0})~e^{-i\omega t} + {\rm c.c.}$
\begin{equation}
C_{\alpha\beta}~(\omega)\equiv 2 \hbar~ {\rm Im}~
({\cal{E}}_{\alpha}~(r_0,~\omega)/p_{\beta});~~~~\omega > 0.
\end{equation}

Note that the dynamical equation (8) can be used to calculate all the 
line shapes (both absorption and emission) for an
anisotropic vacuum by using Eqs. (10), (11), (15) and (16). Note that the quantity in
the bracket in Eq.(16) is the {\em Green's tensor} for the Maxwell  equations.Thus the
procedure for a given geometrical arrangement will consist of evaluation of the
Green's tensor and then the application of (16) to obtain the correlation tensor.

{\bf Interferences in Emission between two Conducting Plates}.

Let us now consider an important problem \cite{vaid} in cavity QED viz the
emission from an atom located between two conducting plates [Fig.2] at $z=0$ and $z=-d$.
The atom is located at $z=-b$. The C's as defined by (4) can be calculated using
(16). These calculations are extremely long. We will only quote the final
result. For this geometry $C_{zx}~,~C_{xz}=0$. Furthermore
the parameters $\gamma's$ and $\kappa's$ entering the master equation (8) can be shown to be 
\begin{equation}
\gamma_i = \gamma_i^{(0)}\left[\Gamma_{\bot}(\omega_i)+\Gamma_{||}
(\omega_i)\right]/2,
\end{equation}
\begin{equation}
\kappa_i = \gamma_i^{(0)}\left[\Gamma_{\bot}(\omega_i)-\Gamma_{||}
(\omega_i)\right]/2,~\gamma_i^{(0)}\equiv\frac{2\omega_i^3 d^2}{3 c^3\hbar} ,
\end{equation}
where
\begin{equation}
\Gamma_{\bot}(\omega)=\frac{3\pi}{2kd} + \frac{3\pi}{kd}\sum_{n=1}^{\cal N}
\left(1 - \frac{\pi^2 n^2}{k^2d^2}\right)\cos^2\left(\frac{b\pi n}{d}\right) ,
\end{equation}
\begin{equation}
\Gamma_{||}(\omega)=\frac{3\pi}{2kd}\sum_{n=1}^{\cal N}\left(1+\frac{\pi^2n^2}
{k^2d^2}\right)\sin^2 \left(\frac{b\pi n}{d}\right), k = \frac{\omega}{c} ,
\end{equation}
and where ${\cal N}$ is the largest integer smaller than $kd/\pi$. The explicit
results (17) and (18) show the expected interferences between the decay
channels. Clearly the interferences will be sensitive to the magnitude of the
magnetic field  which enters through $\omega$. The case of small magnetic fields
is especially interesting. We further note that below the cut off frequency
$k d/\pi~< 1, ~ \Gamma_{||}(\omega)\rightarrow 0,~~~ \Gamma_{\bot}(\omega)
\rightarrow 3\pi / 2 k d$~ and~ $\kappa_i = \gamma_i$.  In this limit quantum
interferences become especially prominent.  
We further note that (a) in the limit $d\rightarrow\infty$, we get
results for emission in the presence of a single conducting plate, (b) the quantities
$\Gamma_{||}$ and $\Gamma_{\bot}$ are related to the emission from a single two
level atom \cite{saran75,knmildow} between the two conducting plates.\\
In conclusion we have demonstrated how the anisotropy of the vacuum field can
lead to new types of interference effects  between the decay of close lying
states. We have related the interference terms to the antinormally ordered
correlation tensor of the vacuum. The anisotropy related interferences are
especially significant for emission from atoms, molecules adsorbed on surfaces
and thus our study opens up the possibility of studying quantum interferences 
in a totally new class of systems.
We have given explicit example of decay
between two conducting plates. We have also shown the role of interference
effects in two photon processes where fluorescence is detected following
excitation by a coherent cw field. Clearly the anisotropy related interference
effects \cite{jpdow} would also be important in considerations of higher order radiative
processes which could be studied in the same manner as two photon processes.

\begin{figure}
\hspace*{3 cm}
\epsfysize 2.5 in
\epsfbox{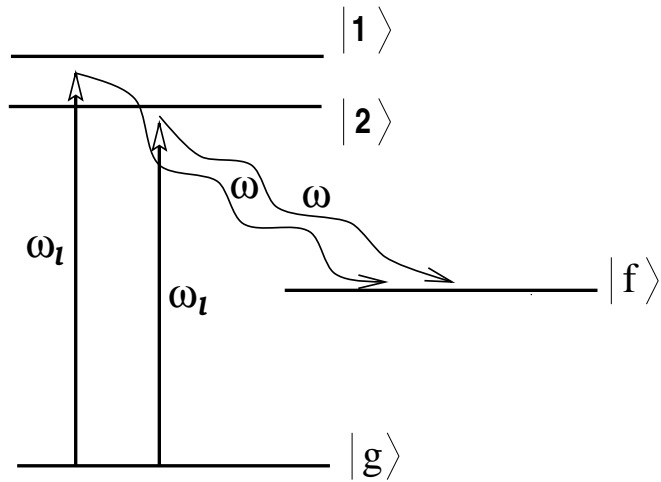}
\vspace*{0.5 cm}
\caption{Schematic diagram of the two interfering pathways contributing to the
two photon process. The wavy arrows represent emission into anisotropic
vacuum.}
\end{figure}

\begin{figure}
\hspace*{3.4 cm}
\epsfysize 3 in
\epsfbox{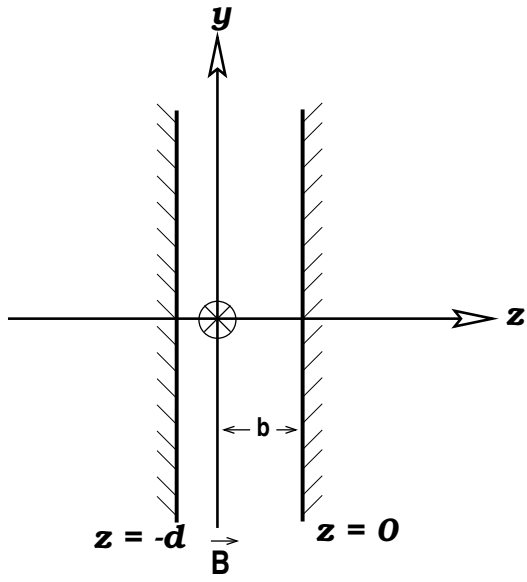}
\vspace*{0.5 cm}
\caption{Emission by an excited three level atom in between two conducting
plates. The magnetic field is along the $y$ direction. The conducting plates lead to
the anisotropy of the vacuum field.}
\end{figure}
\end{document}